\documentclass[aps,twocolumn,superscriptaddress,citeautoscript,showpacs,nobibnotes]{revtex4-1}

\usepackage{float}
\usepackage{amsfonts}
\usepackage{amssymb}
\usepackage{graphicx}
\usepackage{moreverb}
\usepackage{url}
\usepackage{epsf}
\usepackage{color}
\usepackage{multirow}
\usepackage{marginnote}
\usepackage{epstopdf}
\usepackage{ctable}
\usepackage{pgfplots,tikz}

\usepackage[citecolor=blue,citebordercolor=blue,linkbordercolor=green]{hyperref}

\newcounter{mynote}

\newcommand{\siesta}{{\sc Siesta}}  
\newcommand{\abinit}{{\sc Abinit}}  

\begin{document}

\title{Electronic stopping power in a narrow band gap semiconductor from first principles}
\thanks{Published in: \href{http://journals.aps.org/prb/abstract/10.1103/PhysRevB.91.125203}{Phys. Rev. B 91, 125203 (2015)}}

\author{Rafi~Ullah} 
\email[Electronic address: ]{r.ullah@nanogune.eu}
\affiliation{CIC nanoGUNE, Ave. Tolosa 76, 20018 Donostia-San Sebasti\'{a}n, Spain}
\author{Fabiano~Corsetti}
\affiliation{CIC nanoGUNE, Ave. Tolosa 76, 20018 Donostia-San Sebasti\'{a}n, Spain}
\author{ Daniel~S\'anchez-Portal}
\affiliation{Centro de F\'isica de Materiales CSIC-UPV/EHU, 
Paseo Manuel de Lardizabal 5, 20018 Donostia-San Sebasti\'an, Spain}
\affiliation{Donostia International Physics Center, \\
Paseo Manuel de Lardizabal 4, 20018 Donostia-San Sebasti\'an, Spain}
\author{Emilio~Artacho}
\affiliation{CIC nanoGUNE, 20018 Donostia-San Sebasti\'{a}n, Spain}
\affiliation{Donostia International Physics Center, \\
Paseo Manuel de Lardizabal 4, 20018 Donostia-San Sebasti\'an, Spain}
\affiliation{Theory of Condensed Matter, Cavendish Laboratory, University of Cambridge, Cambridge CB3 0HE, United Kingdom}
\affiliation{Basque Foundation for Science Ikerbasque, Bilbao, Spain}

\date{\today}

\begin{abstract}
The direction and impact parameter dependence of electronic stopping power, along with its velocity threshold behavior, is investigated in a prototypical small band gap semiconductor. We calculate the electronic stopping power of H in Ge, a semiconductor with relatively low packing density, using time-evolving time-dependent density-functional theory. The calculations are carried out in channeling conditions with different impact parameters and in different crystal directions, for projectile velocities ranging from 0.05 to 0.6 atomic units. The satisfactory comparison with available experiments supports the results and conclusions beyond experimental reach. The calculated electronic stopping power is found to be different in different crystal directions; however, strong impact parameter dependence is observed only in one of these directions. The distinct velocity threshold observed in experiments is well reproduced, and its non-trivial relation with the band gap follows a perturbation theory argument surprisingly well. This simple model is also successful in explaining why different density functionals give the same threshold even with substantially different band gaps.

\end{abstract}

\maketitle

\section{Introduction}
The study of fast-moving charged particles shooting through solid materials started with Rutherford's famous experiment of showering a gold foil with $\alpha$ particles to substantiate the nuclear model of the atom \cite{Rutherford1911}. Such fast-moving particles strongly perturb the target material. The perturbed state of the medium relaxes either to its original state or a new state with structural defects, depending on the nature of the interaction. The study of such defects, generally referred to as ``radiation damage'', is of great interest from the point of view of applications ranging from nuclear engineering \cite{Gary2007} to biological soft matter for medical applications \cite{Begg2011} and materials engineering for space electronics \cite{Twonsend1994,duzellier}. 

Stopping power is a quantitative measure of the interaction between the projectile and the target medium, defined as the energy transferred from the former to the latter per unit distance traveled through the material. The fast-moving charged particle dissipates its kinetic energy by collisions with the nuclei and the electrons of the medium. Therefore, it is traditional to differentiate between these two distinct dissipation channels; the loss of energy to electronic excitations is known as the electronic stopping power $S_e$, and the loss of energy to the nuclear motion is known as the nuclear stopping power $S_n$.

There is a growing interest in modeling the stopping power of ions with velocities between $0.1$-$1$ atomic units (a.u.\ hereafter) \cite{bohrvel}. In this regime the electronic stopping power (ESP) is  generally dominant; however, at lower velocities the contribution from nuclear collisions also becomes sizable \cite{Mertens2000}. Fermi and Teller\cite {fermi01}, using an electron gas model, found the ESP to be proportional to the projectile velocity for $\mathbf{v}<1$ a.u.\ Lindhard \cite{lindhard} and Ritchie \cite{ritchie}, applying a linear response formalism to an electron gas model of simple metals, predicted a linear velocity dependence within the low projectile velocity limit.  Almbladh \textit{et al.} \cite{almbladh} showed, by calculating the static screening of a proton in an electron gas using density functional theory (DFT), the significant limitations of the linear response treatment. Using DFT Echnique \textit{et al.} \cite{pmech1,pmech02} proposed a full non-linear treatment to account for non-linear effects such as the presence of bound states and the complex electronic structure of the heavy projectiles in the low velocity limit. Recently, the modeling of proton and antiproton stoppings in metals, using jellium clusters as a model of the target, has been extended to intermediate and high projectile velocities using real-time time-dependent density functional (TD-DFT)\cite{tddft,RGross} simulations\cite{borisov, koval}.

Fermi and Teller \cite{fermi01} pointed out that, in case of insulators, the linear velocity dependence of the ESP is only valid in the limit in which the kinetic energy of the projectile is greater than the band gap.  An extensive amount of interesting work has been carried out on the problem of ESP  within the linear response theory \cite{pitarke, pitarkeprb,dorado,juaristi} and non-linear formalism \cite{arista}. A detailed background on the subject can be found  in Ref. \onlinecite{crace} and references therein. A vast majority of these approaches is limited to an electron gas model of metals and do not take into account important features such as the local inhomogeneity of the electron density, core state excitations and band gaps in case of insulators and semiconductors. These features become increasingly important at low velocities. Radiation damage in metals has also been studied, obtaining very interesting 
qualitative results describing the processes in model systems, using explicit electron dynamics within a tight binding model \cite{tightbm01,tightbm02}.

Relatively recently, TD-DFT based first principles calculations of ESP \cite{jpruneda,aacorrea,maz01,maz02} have been performed for insulators and noble metals to explain some interesting effects observed experimentally \cite{Markin2008,Markin2009,Primetzhofer2011,MarkinPRL2009} which do not fit the known theoretical models \cite{crace,pmech1}. These TD-DFT based calculations have successfully reproduced the expected threshold behavior in wide band gap insulators, and the role of \textit{d} electrons in the non-linear behavior found in gold. In contrast, there has not been much work done on semiconductors, except for a study \cite{Hatcher2008} which investigated oscillations in the ESP by varying the atomic number \textit{Z}. However, no systematic velocity-dependent investigation has been attempted at this level of theory. Recent experiments show a possible small velocity threshold for protons in bulk Ge, a system with very small band gap \cite{expHinGe}. The band gap of Ge is almost 20 times smaller than that of LiF while the observed threshold velocity in Ge is only 2 to 3 times smaller. Very little is known about the velocity threshold in small band gap materials.

Experimentally it is almost impossible to measure directly the ESP at velocities $\lesssim 0.2$ a.u., as usually the total stopping power $S=S_n+S_e$ of the medium is measured. The ESP can then be extracted from the measured spectrum using different models \cite{SRIM1985,SRIM2010}. However, a quantitative knowledge of all possible mechanisms contributing to the total stopping power is necessary to extract the electronic component properly. At velocities not much higher than $0.1$ a.u.\ it becomes rather difficult to disentangle the two contributions \cite{Goebl2013}. However, in simulations it is possible to directly access the ESP using TD-DFT based non-adiabatic electron dynamics simulations. In such simulations the projectile is directed along a crystal direction, where it does not get too close to any of the target nuclei. The nuclear contribution to the stopping power, therefore, is negligibly small and can even be completely suppressed by constraining the host atoms to be immobile.

In this study we have investigated the ESP of H in Ge. A small band gap and relatively low packing density makes Ge particularly interesting for the investigation of the threshold behavor which has been observed in wide band gap insulators \cite{MarkinPRL2009,maz02}. 
The simulations have been carried out using an equivalent method to Refs. \onlinecite{maz01, maz02}. Furthermore, we have systematically studied the direction and impact parameter dependence of the ESP, for which very little is known. The accuracy offered by this method, as verified in the satisfactory comparison to experiments below,  allows us to explore these aspects explicitly.
\section{Method}
The calculations are carried out using an extension of the \siesta{} program and method \cite{msiesta,esiesta} which incorporates time-evolving TD-DFT based electron dynamics \cite{danial}. The ground state of the system is calculated with the projectile placed at its initial position. The ground state Kohn-Sham (KS) orbitals \cite{ksdft} serve as initial states. Once the ground state of the system is known, the projectile is given an initial velocity and the KS orbitals are propagated according to the time-dependent KS equation \cite{tddft} using the Crank-Nicholson method with a time step of $1$ as. The forces on the nuclei are muted so that energy is transferred only through inelastic scattering to the electrons. In any case, the projectile velocities are fast enough to  leave little or no time for the nuclei to respond. The projectile velocity itself is similarly kept constant by neglecting forces on the projectile. This allows for a simple extraction of the ESP at a well-defined velocity for each simulation, which is the main aim of our study. The change in velocity, if considered, can be expected to be of no more than $10$\%. 

\begin{figure}[]
\begin{center}
\includegraphics[scale=2]{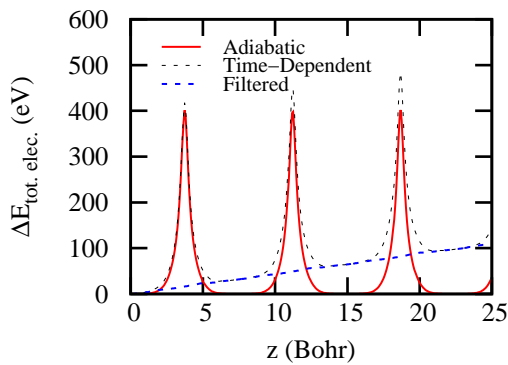}
            \caption{The total energy of the electronic subsystem as a function of the projectile displacement is shown by the dotted (black) line (for a projectile traveling along the $[011]$ direction of Ge at a velocity of $0.6$ a.u.). The solid (red) line shows the adiabatic total energy of the electronic subsystem along the same trajectory. The dashed (blue) line shows the difference between the two, i.e., the non-adiabatic energy contribution.}\label{fig:filter}
\end{center}
\end{figure}

The total energy of the electronic subsystem is recorded as a function of the projectile displacement for a given velocity, as shown by the example in Figure \ref{fig:filter} (dotted black line). The peaks reflect the crystal periodicity. We then adiabatically move the projectile along the same trajectory (i.e., using standard ground-state DFT) and calculate a corresponding adiabatic energy profile (solid red line). Subtracting the adiabatic total electronic energy $E_{a}(z)$ from the time-dependent total electronic energy $E_{td}(z)$ gives an oscillation-free profile of the non-adiabatic energy transfer to the electronic subsystem along the trajectory:
\begin{equation}
\Delta E_{na}(z) =  E_{td}(z)- E_{a}(z);
\end{equation}
$\Delta E_{na}(z)$ is therefore the non-adiabatic contribution shown by the dashed blue line, from which the gradient can easily be extracted by a linear fit; this gives our value for the ESP at that velocity.
\begin{figure}[]
\begin{center}
\includegraphics[scale=0.6]{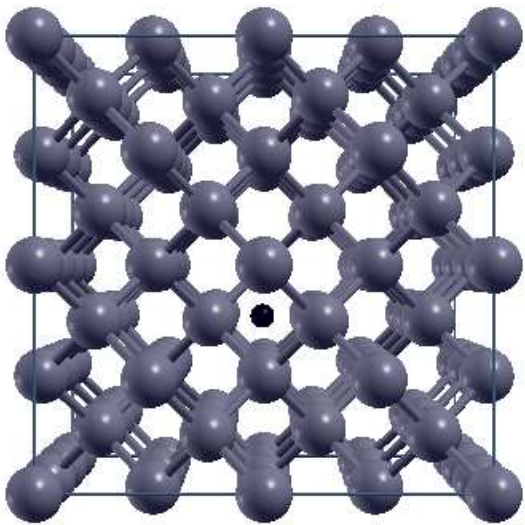}
          \caption{Ge supercell in the [001] direction with H in a channel.}\label{fig:sc}
\end{center}
\end{figure}

The Kohn-Sham orbitals were expanded in a basis of numerical atomic orbitals of finite extent \cite{nao,basis}. A double-$\zeta$ polarized (DZP) basis set was used to represent the valence electrons of the projectile and the host material, while the core electrons were replaced by norm conserving Troullier-Martins pseudopotentials \cite{tmp}, factorized in the separable Kleinman-Bylander (KB) form \cite{kbp}. Pruneda and Artacho \cite{pruneda2004} have studied the validity of pseudopotentials for short range interatomic interactions, showing how the inclusion of core electrons in the valence configuration mitigates the errors from this approximation. Therefore the effect of the Ge pseudopotential was checked by introducing the core (3\textit{d}) electrons into the valence shell, which might be important for the lowest impact parameter trajectories passing very close to some of the Ge ions in the supercell. We did not find a significant error in the ESP for any of the impact parameters shown in our results. Considering the point expected to have the largest pseudopotential error (the lowest impact parameter and the highest projectile velocity), the semicore calculations give an increase of $0.35$ eV/\AA\ (an error of $4$\%). Details of the basis set and the pseudopotentials are given in Appendix \ref{basis}. The sampling of the real-space grid, for representing the electronic density and basis functions for the calculation of some terms of the Hamiltonian matrix \cite{esiesta}, was chosen to correspond to an energy cutoff of 200 Ry. 

A 96-atom supercell (Figure \ref{fig:sc}) constructed by $2\times2\times3$ conventional cubic cells of Ge was used. We have checked the convergence of the ESP with respect to supercell size using a larger 144-atom supercell at a projectile velocity of 0.6 a.u., finding an increase of $0.29$ eV/\AA\ (an error of $4$\%).  A \textit{k}-point mesh of $4\times4\times3$ points generated with the Monkhorst-Pack method \cite{mkpack} corresponding to an effective cutoff length of $22.36$ \AA \cite{moreno} was used after testing its convergence. The exchange and correlation functional was evaluated using the local density approximation (LDA) in the Ceperley-Alder form \cite{LDACA}.

We used the theoretical lattice constant, which was found to be $5.59$ \AA, compared to an experimental value of $5.66$ \AA. This underestimation of $\sim1$\% is typical for the LDA. An indirect band gap of $0.70$ eV was found for bulk Ge, compared with an experimental value of $0.74$ eV (at $0$ K). However, it is important to note that this good agreement is fortuitous, as DFT with LDA generally either underestimates the band gap or does not produce one at all. Pseudopotential can be one of the sources of cancellation of errors \cite{pawGe} along with a smaller lattice parameter which tend to open the band gap.  Lee \textit{et al.} \cite{cheliko2}, using a plane-wave method, have reported an indirect band gap of $0.41$ eV. Much larger band gap, up to $0.81$ eV \cite{cheliko}, have been reported depending upon the details of the calculation. The dependence on the density functional was checked by repeating the calculations for the Perdew-Burke-Ernzerhof (PBE) functional \cite{PBE01}, for which the theoretical lattice constant was found to be $5.78$ \AA~with a direct band gap of $0.33$ eV.

 In order to check the convergence of our basis in \siesta{}, we have also computed the band structure with the plane-wave DFT code \abinit{} \cite{abinit}, making use of exactly the same pseudopotential including the same choice of local potential and KB projectors, and a high kinetic energy cutoff of 95 Ry for the basis. The agreement for the valence and low-lying conduction bands is excellent, although we find a slightly smaller band gap of $0.58$ eV with the plane-wave calculation (see Appendix \ref{abinitband}).

The projectile trajectories are chosen along the $[001]$, $[011]$, and $[111]$ directions. A sectional view of the simulation box orthogonal to the [001] channel is shown in Figure \ref{fig:sc}. Different representative impact parameters are considered within the $[001]$, $[011]$, and $[111]$ channels. The projectile velocities range from $0.05$ a.u.\ to $0.6$ a.u.\ for each trajectory.

\section{Results and Discussion}

\begin{figure}[]
\includegraphics[scale=2.5]{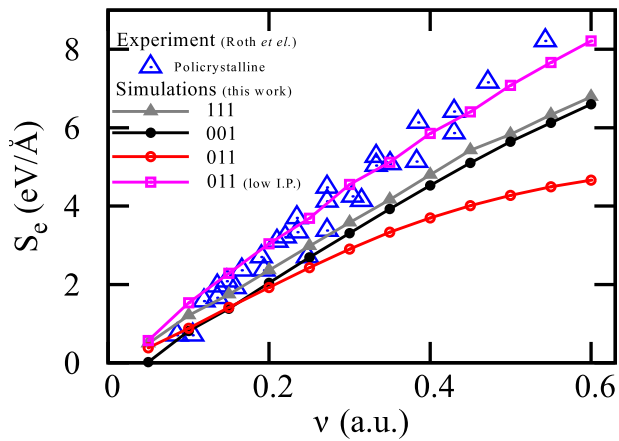}
\caption{Electronic stopping power ($S_e$) vs velocity ($v$) of a H projectile in bulk Ge along different crystal directions, as obtained from TD-DFT, and  compared with the experimental measurements (empty triangle dat points) reported in Ref. \cite{expHinGe}. The trajectories in all the three directions are along the centers of respective channels with one additional trajectory in the [011] direction (empty square data points) at a very low impact parameter, ($0.24$ Bohr position 1 in Figure \ref{fig:imp011}).} 
\label{sp:expthe}

\end{figure}

In an experiment with a polycrystalline sample the projectile gets channeled along different crystal directions. We have therefore taken into account the direction and impact parameter dependence.  We have computed the ESP along three different channels. The calculated ESP is compared with experimentally measured values by Roth \textit{et al.} \cite{expHinGe} in Figure \ref{sp:expthe}. 

\subsection{The velocity threshold}
The ESP varies linearly with projectile velocity, intercepting zero at a finite velocity. This indicates a definitive threshold. Roth \textit{et al.} \cite{expHinGe} determine the threshold velocity, by extrapolating the experimental data, to be $0.027$ a.u.\ $\pm10$\%. We have found the threshold velocity to be different for different channels. It is $0.05$ a.u.\ in the [001] direction and $0.03$ a.u.\ in the [111] and [011] directions.

The threshold behavior has been observed in insulators both experimentally and theoretically. From perturbation theory a relationship between the projectile velocity and electronic transitions is given by (see, {\em e.g.,} Ref. \onlinecite{artacho})
\begin{equation}
\mathbf{v}_{th}\cdot\Delta\mathbf{k}=\varepsilon_g,
\label{excit}
\end{equation}
where $\mathbf{v}$ is the projectile velocity, $\Delta\mathbf{k}$ is the change in momentum in electronic excitations, and $\varepsilon_g$ is the band gap and we are taking $\hbar=1$ for simplification through out this article. This relation can be deduced by requiring the conservation of energy and momentum for a two-particle collision event in the limit of mass of projectile $M\to\infty$ (see Appendix \ref{conser}). Following equation \ref{excit}, the velocity threshold for an indirect band gap modelled as in Figure
\ref{inband} would correspond to the relation (see Appendix \ref{conser})
\begin{equation}
\varepsilon_g = {1 \over 2} (m_e + m_h) {v}^2_{th} + {k}_0{v}_{th} ,
\label{eq:indbg}
\end{equation}
where $m_e$ and $m_h$ are the electron and hole masses,
respectively, $k_0$ is the difference in crystal momentum between the valence band maximum and the conduction band minimum, and $\varepsilon_g$ is the indirect band gap. It follows that for small $k_0$ the threshold returns to
the direct band gap behaviour (see Ref. \onlinecite{artacho}), and $v_{th} \propto \sqrt{\varepsilon_g}$.
In the case when both parabolas are thin on the scale of $k_0$,
{\em i.e.}, when $k_0 \gg \sqrt{(m_e+m_h) \varepsilon_g}$, the threshold 
velocity rather goes as $v_{th} = {\varepsilon_g \over k_0} \,$
and is thus linear with $\varepsilon_g$.

\begin{figure}[]
\begin{center}
\includegraphics[scale=2.0]{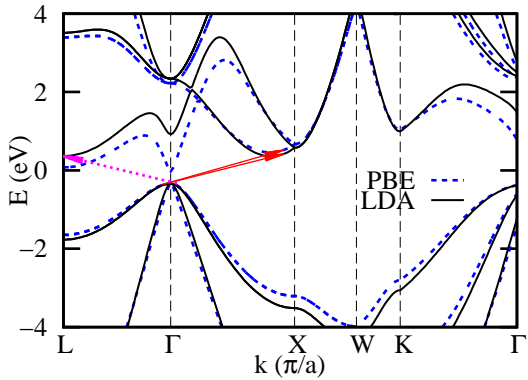}
\caption{Band structure of bulk Ge, calculated using PBE (dashed blue line) and LDA (solid black line). The valence band maxima from the two calculations are aligned with each other for clarity. The two solid (red) arrows illustrate the threshold velocity corresponding to electron-hole excitations in both cases following equation \ref{excit} in the [001] direction. The dotted (magenta) arrow shows the same (LDA only) in the [111] direction.}
\label{fig:ldapbeband}
\end{center}
\end{figure}

This argument implies that, for parabolic bands, below a threshold velocity the ESP would drop to zero. For the case of periodic bands, however, this threshold would not be strict, but can still be defined within some accuracy depending on the smoothness of the projectile's potential convoluted with the relevant electronic wave functions \cite{artacho}. From Equation \ref{excit}, a threshold velocity in a given direction can be estimated from the band structure of the material by finding the gradient of the line which is a joint tangent to the valence and conduction bands, shown by the arrow in Figure \ref{inband}. The threshold velocity estimated from the band structure in the [001] direction is found to be $0.053$ a.u.\ as shown in Figure \ref{fig:ldapbeband} (solid arrows), which is in good agreement with the calculated value of $0.05$ a.u. in the same direction. Furthermore, the reason for finding different threshold velocities in different directions becomes clear, as the gradient of the joint tangent line in the [111] direction (dotted arrow in Figure \ref{fig:ldapbeband}) is different and smaller, in qualitative agreement with the TD-DFT calculations. Although the mentioned experiments average out this direction dependence, here we can relate it with the band structure of the host material.
\begin{figure}[H]
\begin{center}
\includegraphics[scale=1]{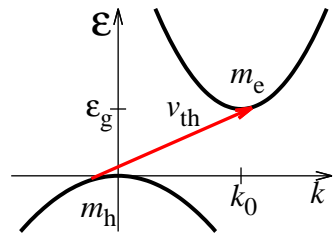}
\caption{Schematic illustration of the relationship between an indirect band gap and the threshold velocity. The arrow shows a common tangent line from the top valence band to the bottom of conduction band.}
\label{inband}
\end{center}
\end{figure}
\begin{figure}[h]
\begin{center}
\includegraphics[scale=2.0]{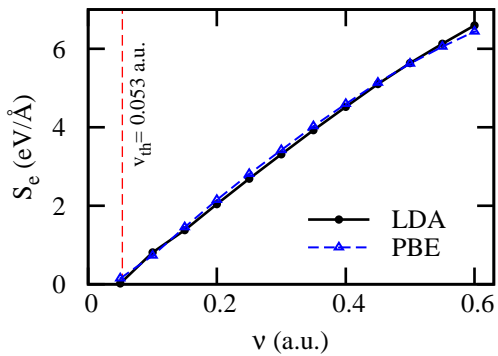}
\caption{The ESP, calculated using the PBE (dashed blue line with triangle data points) and LDA (solid black line with circle data points) functionals, in the [001] direction. The dashed (red) line shows the threshold velocity estimated from the band structure.}
\label{fig:ldavspbe}
\end{center}
\end{figure}
The comparison between LDA and PBE results in Figures \ref{fig:ldapbeband} and \ref{fig:ldavspbe} is of special interest. The electronic band gaps differ by a factor of 2, and yet the ESP shows no significant difference. The LDA functional produces an indirect band gap of $0.70$ eV, while the PBE functional produces a direct band gap of $0.33$ eV. The calculated band structures are shown in Figure \ref{fig:ldapbeband}. However, the ESP calculated using LDA and PBE does not differ significantly at low velocities, and the two calculations produce almost the same threshold. This is a clear indication that the threshold phenomenon is not straightforwardly related to the band gap. The gradient of the joint tangent line of the valence and conduction bands in both cases is almost the same (shown by the solid arrows in Figure \ref{fig:ldapbeband}). This suggests that the behavior of the ESP threshold at low velocities is rather related to the indirect band gap in the given direction regardless of its being the absolute gap. This further supports the above described model of the ESP threshold. The fact that the relation  in Equation \ref{eq:indbg} is accurate using the unperturbed host band structure is somewhat surprising. Such agreement is due to the fact that the perturbing projectile potential does not significantly affect the band structure around the gap.

\subsection{Direction and impact parameter dependence}

\begin{figure}[]
\begin{center}
\includegraphics[scale=0.85]{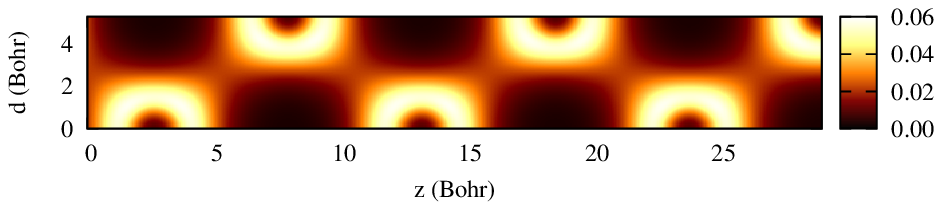}
\includegraphics[scale=0.85]{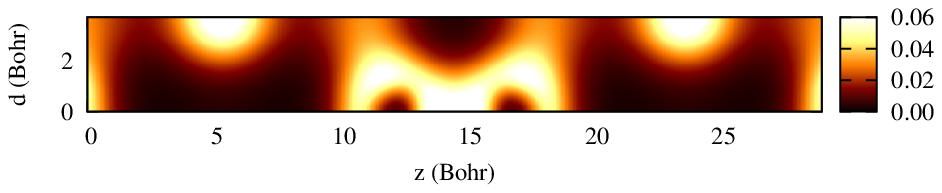}
\includegraphics[scale=0.85]{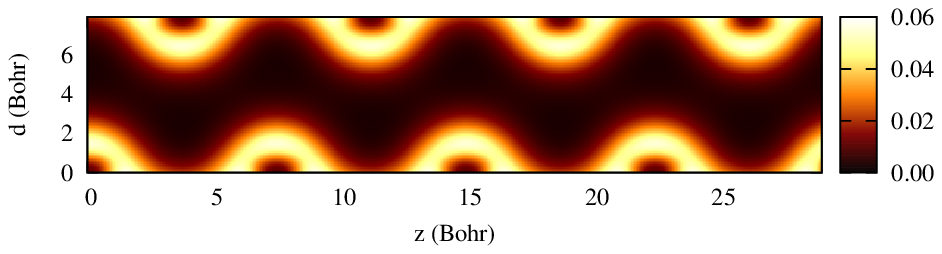}
\caption{The projected electronic densities along the trajectories of projectile in different channels, top [001], middle [111], bottom [011]. The depicted planes are defined by the projectile direction of propagation ($z$) and a high symmetry perpendicular direction $d$ (the [$011$] in case of the [$001$] channel). The electron density increases from dark to bright.  }
\label{dens}
\end{center}
\end{figure}

\begin{figure}[]
\begin{center}
\includegraphics[scale=2]{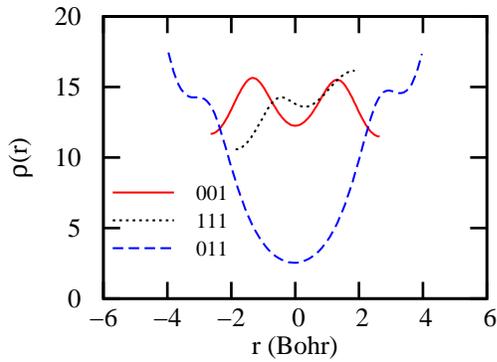}
            \caption{The projected density is averaged over the $z$-axis for all three channels.}\label{fig:avdens}
\end{center}
\end{figure}

We have found that the ESP strongly depends on direction in the crystal, particularly at high velocities. The difference in the ESP between the [111] and [001] channels is up to $3$\%, and between these two and the [011] channel it is up to $33$\%. The electron density along these channels is shown in Figure \ref{dens} in suitable planes. The electron density is then averaged over the $z$-axis, as shown in Figure \ref{fig:avdens}. The direction with the lowest ESP for a channeled projectile ($[011]$) has a lower average density in the center of the channel compared with the two other channels. For $[001]$ and $[111]$ the averaged density is not significantly different, similarly what happens for the ESP. This suggests that the ESP in channeling conditions can be related to the average density along the trajectory, corroborating and supporting assumptions and approximations used in the literature \cite{Nagi01, Nagi02, juaristi2003, juaristi2013}.

\begin{figure}[]
\begin{center}
\includegraphics[scale=2]{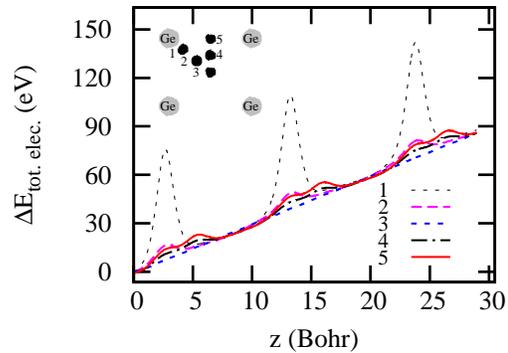}
            \caption{Electronic energy against distance along the different projectile trajectories in the [001] direction. The projectile velocity for all the trajectories is $0.5$ a.u.. The inset shows a sectional view of the [001] channel and the trajectories. The gray circles represent Ge atoms in different transverse planes (defining the channel), while the black circles show the projectile positions for different impact parameters.}\label{fig:imp}
\end{center}
\end{figure}

\begin{figure}[]
\begin{center}
\includegraphics[scale=2]{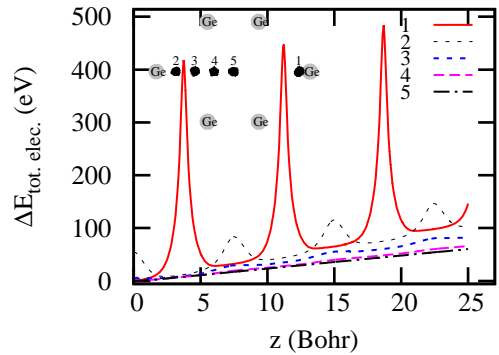}
            \caption{Electronic energy against distance along the different projectile trajectories in the [011] direction. The projectile velocity for all the trajectories is $0.5$ a.u.. The inset shows a sectional view of the [011] channel and the trajectories. The gray circles represent Ge atoms in different transverse planes (defining the channel), while the black circles show the projectile positions for different impact parameters.}\label{fig:imp011}
\end{center}
\end{figure}

We have simulated five different trajectories in the [001] channel, as shown in the inset of Figure \ref{fig:imp}. The five trajectories are chosen to sample different impact parameters (different closest distance to any of the host atoms) within the channel. For each trajectory we show the total energy of the electronic subsystem versus distance for a given velocity of $0.5$ a.u. in Figures \ref{fig:imp} and \ref{fig:imp011}. The plots in Figure \ref{fig:imp} show the energy profile along the [$001$] channel; the periodic variation in the electronic energy reflects the periodicity of the crystal. A larger variation is seen for the trajectories with the lowest impact parameters, as should be expected; however, the base-lines of all the trajectories have the same gradient, which shows that, in this direction, the ESP is quite insensitive to impact parameters.  A similar calculation in the [111] direction gives the same result (not shown). However, the ESP strongly depends on impact parameter in the [011] direction. The total electronic energy profile for five different trajectories in this direction is shown in Figure \ref{fig:imp011}. The change in ESP from the highest impact parameter, i.e., the center of channel (empty circle data points in Figure \ref{sp:expthe}) and the lowest impact parameter, i.e., close to the edge of channel (empty square data pionts in Figure \ref{sp:expthe}) changes by  a factor of 2. Again looking at the average density in the $[011]$ direction (Figure \ref{fig:avdens}), we can see that it changes by a factor of 3 from the center to the edge of the channel. This reflects the proposed strong correlation between the ESP and the averaged local density within a small radius of the impact parameter. It is to be expected that such a radius (or cross section) would increase for slower projectiles. This is verified by the larger slope of the ESP for the center of the [$011$] channel trajectory for lower velocities. Indeed, the low velocity limit displays the same behavior for all trajectories, indicating that the larger cross section is seeing the same average electron density in all the cases.

In experiment the ESP is naturally averaged over different directions and impact parameters, and precise knowledge of this averaging mechanism would be necessary to obtain a comparable average from our calculations. We have not attempted to do so, although it is clear from Figure \ref{sp:expthe} that any such averaging would result in a slight underestimation with respect to experiment, especially for high velocities.


\section{Summary}
We have systematically studied the different aspects of the ESP of H in bulk Ge, a representative narrow band gap semiconductor for which good experimental results are available. We have learned that the ESP is sensitive to the crystal direction and, in certain directions, to the choice of impact parameter. A detailed model is needed to average the calculated ESP over different directions. Similarly to what is known for insulators, a finite velocity threshold is found in the calculations, in agreement with what has been observed experimentally. Here the threshold is found to be much better defined (a strict threshold) than in previous similar studies of the ESP of H in LiF \cite{jpruneda}, a wide band gap insulator. Careful analysis of the band structure of bulk Ge indicates that the threshold phenomenon is connected to the indirect band gap in given crystal directions. Our results give further insight into the understanding of the threshold behavior of the ESP in materials with a band gap.

\begin{acknowledgments}
We are thankful to M. A. Zeb, A. Arnau, J. I. Juaristi, J. M. Pitarke, P. Bauer, D. Roth, and A. Correa for useful discussions. The financial support from MINECO-Spain through Plan Nacional Grant No.\ FIS2012-37549-C05-01, FPI Ph.D. Fellowship Grant No.\ BES-2013-063728, and Grant No.\ MAT2013-46593-C6-2-P along with the EU Grant ``ElectronStopping" in the Marie Curie CIG Program is duly acknowledged. SGIker (UPV/EHU, MICINN, GV/EJ, ERDF and ESF) support is gratefully
acknowledged.
\end{acknowledgments}
\appendix
\section{}\label{basis}
The parameters needed for the generation of the basis set used in this work, according to the procedure explained in Ref. \onlinecite{nao}, are given in Table \ref{dzp}. The parameters need to generate the pseudopotentials  are listed in Table \ref{pseudo}.
\subsection{Basis Set}
\begin{table}[H]
\caption{Cutoff radii $r(\zeta_1)$, $r(\zeta_2)$ of first and second zeta functions respectively, and the soft-confinement potential's internal radius $r_i$ are in Bohr; the soft-confinement potential pre-factor $V_0$ is in Ry.}
\begin{center}
\begin{tabular}{l@{\hskip 0.3in}*{2}{c@{\hskip 0.3in}}*{4}c}

\hline
Species & {\textit n }& {\textit l} &$V_0$& $r_i$&$r(\zeta_1)$ & $r(\zeta_2)$\\
\specialrule{.1em}{.1em}{0.1em}
Ge & 3 & 2 & 50&6&6.50 & \\
 & 4 & 0&50&6 & 6.50 & 5.00 \\
& 4 & 1 &50&6& 6.50 & 4.50 \\
& 4 & 2 &50&6& 6.50 & \\
\\
H & 1 & 0 & 50&6&7.00 & 2.90 \\
&2&1&1000&0&6.00&\\
\hline
\end{tabular}

\label{dzp}
\end{center}
\end{table}

\subsection{Pseudopotential}

\begin{table}[H]
\caption{Matching radii for each of the angular momentum channels of Ge and H. All lengths are in Bohr.}
\begin{center}
\begin{tabular}{l@{\hskip 0.3in}*{3}{c@{\hskip 0.3in}}c}

\hline
Species & {\textit s }& {\textit p} & ${\textit d}$ & ${\textit f}$ \\
\specialrule{0.1em}{0.1em}{0.1em}
Ge($4s^24p^2$) & 2.06& 2.85 & 2.58 & 2.58 \\
Ge($3d^{10}4s^24p^2$) & 1.98& 1.98 & 1.49 & 1.98 \\
H($1s^2$) & 1.25 & 1.25 & 1.25 & 1.25 \\
\hline
\end{tabular}

\label{pseudo}
\end{center}
\end{table}
\section{}\label{abinitband}
\begin{figure}[H]
\begin{center}
\includegraphics[scale=2]{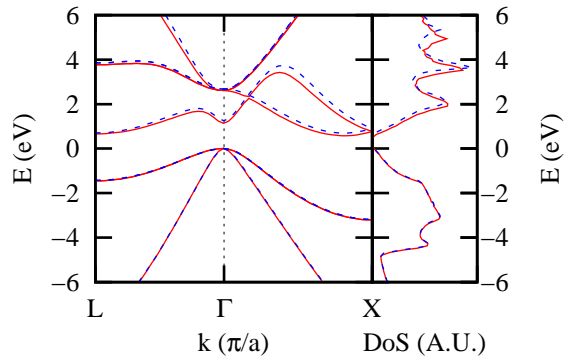}
            \caption{The solid and dashed lines represent \abinit\ and \siesta\ calculations, respectively. }\label{fig:abinitsies}
\end{center}
\end{figure}

The band structure and density of states of bulk Ge calculated using \siesta\ (LCAO) and \abinit\ (Plane Waves) is compared in Figure \ref{fig:abinitsies}. The same pseudopotential (and its local and non-local components) is used in both codes.
\section{}\label{conser}
\subsection{Threshold Velocity}
This is a known relationship that can be obtained in several different ways; here, we present one such way of deriving it.
If a particle of mass $m$ and initial momentum $\mathbf{k}_i$ collides with another particle of mass $M$ and initial momentum $\mathbf{K}_i$, conservation of momentum requires that
\begin{equation}
\Delta \mathbf{k} \equiv \mathbf{k}_f - \mathbf{k}_i = \mathbf{K}_i-\mathbf{K}_f,
\label{momcons}
\end{equation}
where $\mathbf{k}_f$ and $\mathbf{K}_f$ are the final momenta of the particles, respectively, and $\Delta\mathbf{k}$ denotes the change in momentum. Conservation of energy requires that
\begin{equation}
\varepsilon_f - \varepsilon_i = \frac{1}{2M}({K}_i^2-{K}_f^2),
\label{enercons}
\end{equation}
where $\varepsilon_i$ and $\varepsilon_f$ are initial and final energies of the particle of mass $m$, respectively. From equation \ref{momcons}, we can write
\begin{equation}
{K}_i^2-{K}_f^2=2\Delta \mathbf{k}\cdot\mathbf{K}_i-\Delta{k}^2.
\label{subs}
\end{equation}
On substituting equation \ref{subs} in equation \ref{enercons}, we obtain
\begin{equation}
\varepsilon_f - \varepsilon_i = \frac{1}{M}\mathbf{K}_i\cdot\Delta\mathbf{k}-\frac{1}{2M}\Delta{k}^2.
\label{sub2}
\end{equation}
In the limit $M\to\infty$, the second term in equation \ref{sub2} vanishes, and the rest simplifies to
\begin{equation}
\varepsilon_f - \varepsilon_i = \mathbf{v}\cdot\Delta\mathbf{k},
\label{direc1}
\end{equation}
where $\mathbf{v}=\frac{\mathbf{K}_i}{M}$. The smallest excitation in the system would require $\varepsilon_f - \varepsilon_i=\varepsilon_g$, where $\varepsilon_g$ is the band gap of the material, with an accompanying change in momentum $\Delta\mathbf{k}$ of the electron undergoing the transition. The threshold velocity of the projectile at the onset of energy loss would therefore relate to the band gap as:
\begin{equation}
\varepsilon_g = \mathbf{v}_{th}\cdot\Delta\mathbf{k}.
\label{direc}
\end{equation}
\subsection{Indirect band gap}
The argument for deducing the excitation condition in a direct band gap case can be extended to the case of parabolic bands with an indirect band gap. The condition for the direct band gap [$\varepsilon_g=\frac{1}{2}(m_e+m_h){v}_{th}$] can be found in Ref. \onlinecite{artacho}. A geometrical way to proceed for the indirect band gap is to find the conditions for which a straight line (corresponding to the red arrow in Figure \ref{inband}) would cross both of the parabolas, and from these derive the limiting velocity value below which there is no crossing. Considering first the parabola for electrons, we can write
\begin{equation}
\varepsilon_e=\frac{1}{2m_e}|\mathbf{k}_e-\mathbf{k}_0|^2+\varepsilon_g.
\end{equation}
The transition line $\varepsilon_t=\mathbf{k}_e\cdot\mathbf{v}+\varepsilon_0$ should cross the                                                                             parabola $\varepsilon_e$, where $\varepsilon_0$ is a constant
defining the vertical positioning of the transition line of slope $\mathbf{v}$ 
(red arrow in Figure \ref{inband}):
\begin{equation}
\frac{1}{2m_e}|\mathbf{k}_e-\mathbf{k}_0|^2+\varepsilon_g=\mathbf{k}_e\cdot\mathbf{v}+\varepsilon_0.
\label{eq:etrans}
\end{equation}
Here for simplicity we consider that $\mathbf{k}_0$ and $\mathbf{v}$ are collinear. Furthermore, since we are interested in obtaining an equation for the threshold velocity, we can consider that $\mathbf{k}_{e}$ is parallel to $\mathbf{v}$ without loss of generality. The equation \ref{eq:etrans} is quadratic in ${k}_e$ and can be solved to give
\begin{widetext}
\begin{equation}
{k}_e={k}_0+m_e{v}\pm\sqrt{({k}_0+m_e{v})^2-2m_e(\varepsilon_g-\varepsilon_0)-{k}_0^2}
\label{eq:esol}.
\end{equation}
\end{widetext}
Similarly, for holes we can write
\begin{equation}
\varepsilon_h=-\frac{{k}_h^2}{2m_h}.
\end{equation}
Again, the transition line $\varepsilon_t=\mathbf{k}_h\cdot\mathbf{v}+\varepsilon_0$ should cross this parabola. Equating the two gives a quadratic equation in ${k}_h$ which can be solved to give
\begin{equation}
{k}_h=-{m_h}{v}\pm\sqrt{({m_h}{v})^2-2m_h\varepsilon_0}.
\label{eq:hsol}
\end{equation}
The two conditions \ref{eq:esol} and \ref{eq:hsol} (for electrons and holes, respectively) can be combined as
\begin{equation}
\frac{1}{2}m_h{v}^2\ge\varepsilon_0 \ge\varepsilon_g-\frac{1}{2}m_e{v}^2-{k}_0{v};
\label{eq:econd}
\end{equation}
for that to be possible,

\begin{equation}
\frac{1}{2}m_h{v}^2\ge\varepsilon_g-\frac{1}{2}m_e{v}^2-{k}_0{v},
\label{eq:ehcond}
\end{equation}
leading to
\begin{equation}
\varepsilon_g\le \frac{1}{2}(m_e+m_h){v}^2+{k_0}{v},
\end{equation}
or, at ${v}={v}_{th}$,
\begin{equation}
\varepsilon_g= \frac{1}{2}(m_e+m_h){v}^2_{th}+{k_0}{v}_{th}.
\end{equation}


%

\end{document}